\documentclass[draft]{agujournal2019}
\usepackage{url} 
\usepackage{lineno}
\usepackage[inline]{trackchanges} 
\usepackage{soul}
\drafttrue

\journalname{JGR: Atmospheres}

\begin{document}

%
%


\title{Reconciling observations of solar irradiance variability with cloud size distributions}

%
%




\authors{Wouter B. Mol\affil{1}, Bart J. H. van Stratum\affil{1}, Wouter H. Knap\affil{2}, Chiel C. van Heerwaarden\affil{1}}

\affiliation{1}{Meteorology and Air Quality Group, Wageningen University, Wageningen, The Netherlands}
\affiliation{2}{Royal Netherlands Meteorological Institute, De Bilt, The Netherlands}






\correspondingauthor{Wouter Mol}{wbmol@wur.nl}




\begin{keypoints}
\item Surface solar irradiance variability scales with a power law over multiple orders of magnitude, set by the distribution of cloud sizes
\item Broken and scattered boundary layer clouds cause most of the intra-day solar irradiance variability
\item Cloud shadow size can be constructed from cloud size by accounting for cloud edge transparency with an approximate length scale of 25 meters
\end{keypoints}

%
%

%
%


\begin{abstract} 
Clouds cast shadows on the surface and locally enhance solar irradiance by absorbing and scattering sunlight, resulting in fast and large solar irradiance fluctuations on the surface.
Typical spatiotemporal scales and driving mechanisms of this intra-day irradiance variability are not well known, hence even one day ahead forecasts of variability are inaccurate.
Here we use long term, high frequency solar irradiance observations combined with satellite imagery, numerical simulations, and conceptual modelling to show how irradiance variability is linked to the cloud size distribution.
Cloud shadow sizes are distributed according to a power law over multiple orders of magnitude, deviating only from the cloud size distribution due to cloud edge transparency at scales below $\approx$ 750 meters.
Locally cloud-enhanced irradiance occurs as frequently as shadows, and is similarly driven mostly by boundary layer clouds, but distributed over a smaller range of scales. 
We reconcile studies of solar irradiance variability with those on clouds, which brings fundamental understanding to what drives irradiance variability.
Our findings have implications for not only for weather and climate modelling, but also for solar energy and photosynthesis by vegetation, where detailed knowledge of surface solar irradiance is essential.
\end{abstract}

\section*{Plain Language Summary}
The amount of sunlight received at a given location on the Earth's surface varies on many temporal and spatial scales.
Clouds have a large influence on the daily and local scale by creating alternating patterns of shadows and extra bright sunlight, which can fluctuate back and forth multiple times per minute.
Weather and climate models do not capture these details, because they are necessarily simplified in order to save computational time.
We aim to learn what the typical time and spatial scales of these shadows and sunny patterns are, and where they originate from.
Our study is based on ten years of detailed sunlight observations, satellite imagery of clouds, and cloud-resolving atmospheric modelling. 
We find that horizontal scales in sunlight variability are almost identical to those of clouds. 
Only at the smallest scales, where clouds are so small that they become transparent, do the observations of sunlight variability deviate from clouds sizes.
These observations can be reproduced with a simple model based on the size distributions of clouds.
This knowledge could help guide those who try to include sunlight variability in a simple way into their analysis and modelling of, for example, solar energy and photosynthesis.

%
%

\section{Introduction}
Surface solar irradiance can exceed clear-sky, even extraterrestrial irradiance \cite<e.g.>{yordanov_extreme_2015, gueymard_cloud_2017}, caused by the scattering of radiation by broken clouds, referred to as cloud enhancement. 
Simultaneously these clouds cast shadows, creating ever changing and moving patterns of alternating high and low irradiance at the surface, resulting in high ramp rates and heterogeneity in surface heat fluxes.
The resulting spatiotemporal variability on the intra-day and local scale poses a challenge for stable integration of solar energy in the electricity grid by being unpredictable \cite{liang_emerging_2017, yang_review_2022}, while society will increasingly depend on solar energy in mitigating climate change \cite{IPCC_2022_WGIII_Ch_6}.
Furthermore, the global carbon, water, and energy cycles are affected by heterogeneous distribution of solar irradiance caused by clouds \cite{mercado_impact_2009, lohou_surface_2014, keenan_widespread_2019, humphrey_soil_2021, hogan_entrapment_2019}.
Radiative transfer is well-described in theory, yet necessarily simplified in model implementations due to its high computational demand \cite<e.g.>{hogan_flexible_2018}.
One such simplification is the two-stream approach, which calculates radiative transfer in only the up and downward direction, whereas in reality, photons travel in all directions.
This means that even atmospheric models with fully resolved cloud fields are unable to resolve realistic surface irradiance variability.
Apart from directly negative consequences for solar energy applications, these errors become more pronounced the better such models resolve individual clouds, and can feed back to other components of the model.
Most notably, heterogeneity in surface irradiance can feed back through surface heat fluxes to alter the cloud field and domain-mean irradiance \cite{lohou_surface_2014, jakub_role_2017, veerman_case_2022}.
Given the aforementioned importance of accurately resolved or forecast irradiance variability, it is imperative to understand on more fundamental level how irradiance variability is characterized and where it originates from.

Previous analyses of cloud shadow and enhancement duration have shown various statistical distributions that suggest irradiance variability occurs over a wide range of scales \cite{gu_cloud_2001, tomson_fast_2010, tabar_kolmogorov_2014, lappalainen_analysis_2016, madhavan_multiresolution_2017, kivalov_quantifying_2018, jarvela_characteristics_2020}, but there is no consensus on how to characterise the distributions.
This, in part, is owed to the studies focusing on different aspects of surface irradiance, using different instrumental setups, and sampling in different climatological regions and times of year. 
The debate has also been limited by the often lacking multi-year irradiance observations (for statistical convergence) sampled with at least 1 Hz temporal resolution, necessary to capture the details of the fast fluctuations \cite{yordanov_100-millisecond_2013}.
Examples most relevant to this study include \citeA{gu_cloud_2001}, who identify a 5/3 power law slope in irradiance power spectra below 10 minutes as turbulence-like, based on three weeks of 1-minute resolution observations.
Based on a year of 1 Hz resolution data, this was also shown by \citeA{tabar_kolmogorov_2014} for a range of 10 to 1000 seconds, and they suggest this is rooted in the cloud size distribution, which similarly scales with $\approx$ 5/3 \cite{wood_distribution_2011}.
Based on a dense spatial network of 99 pyranometers sampling at 1 Hz, \citeA{madhavan_multiresolution_2017} find for overcast cases similar scaling in spatial power spectra between 50 and 1500 meters, but exponents vary for larger spatial scales or different cloud cover types.
Evidence for a link to cloud size distributions in these studies is empirical and limited by the data resolution or temporal range of their observations, or the connection between temporal irradiance data and spatial cloud sizes is not made.
However, the irradiance variability power law scaling and link to cloud sizes is certainly an appealing idea that deserves a more thorough analysis, which we will provide with a more robust dataset in the present study. 

The Baseline Surface Radiation Network \cite<BSRN,>{driemel_baseline_2018} station of the meteorological observatory in Cabauw, the Netherlands, has been measuring solar irradiance every second since 2005, with high standards of maintenance and quality control.
The long measurement record, resolution, and multi-component properties of this dataset enable the analysis of the climatology of intra-day irradiance variability in meteorological context.
Details of the data and methodology are described further in Section \ref{sec:method}.
From these observations, we describe the typical spatiotemporal scales of surface irradiance variability, how these vary by cloud type, and how variability is linked to cloud size distributions (Sections \ref{sec:dists} to \ref{sec:powerlaw}).
We furthermore provide an interpretation of this link by doing an sensitivity analysis (Section \ref{sec:expsense}), and by running a cloud resolving model that is able to reproduce observations (Section \ref{sec:resmodel}).

\section{Data and Methods}\label{sec:method}
Our analysis makes extensive use of observational data, which is described in Section \ref{sec:data}. 
The analysis itself starts with irradiance time series classification into cloud shadow and enhancement categories, the methodology of which is described in Section \ref{sec:methodclass}.
From these processed time series, we construct distributions of the spatiotemporal scales of cloud shadows and enhancements, and are further interpreted using cloud observations.
In order to make the link between these distributions and cloud sizes, we make use of distribution fitting and a conceptual model, laid out in Section \ref{sec:fitmodel}.
For further interpretation, we use a cloud resolving model and an idealized cloud model, both explained in Section \ref{sec:nummodel}.

\subsection{Observational Data}\label{sec:data}
\begin{figure}[ht]
\centering
\includegraphics[width=1\textwidth]{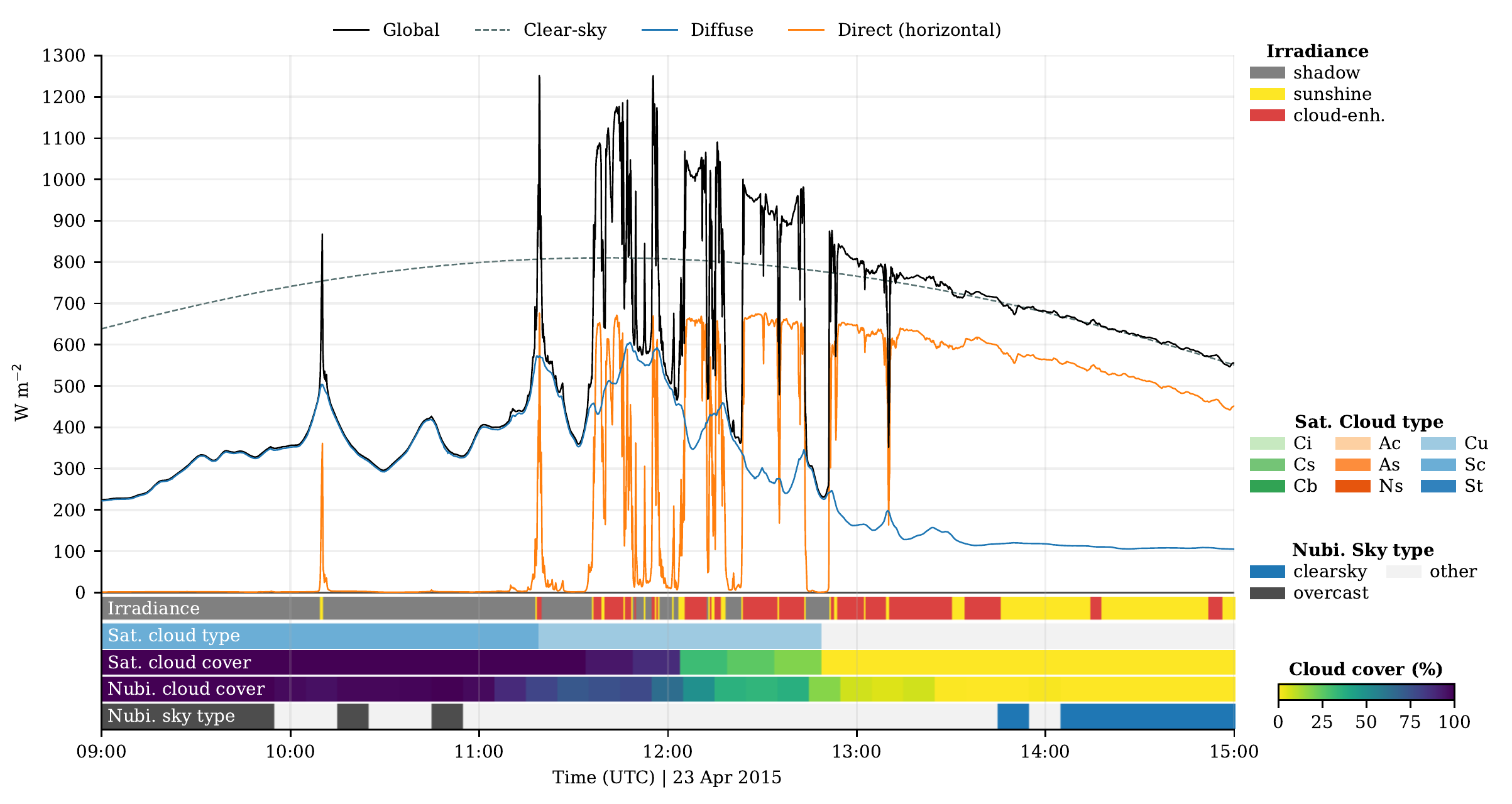}
\caption{\textbf{Cabauw BSRN time series, irradiance classifications, and cloud observations} for part of 23 April 2015, which features a weather transition from overcast to clear-sky via a period of frequent and strong cloud enhancements. The three measured irradiance components, global, diffuse, and direct horizontal irradiance are shown together with modelled (CAMS McClear) clear-sky irradiance. The satellite and nubiscope observations show an overcast stratocumulus (Sc) field dissolving via cumulus (Cu) into to clear-sky conditions.}
\label{fig:example_timeseries}
\end{figure}

The BSRN station of the atmospheric observation site in Cabauw, the Netherlands (\url{https://ruisdael-observatory.nl/cabauw/}) has logged irradiance measurements at a resolution of 1 second from 2005 onward. 
Direct, diffuse, and global horizontal irradiance for a subset of 10 years (2011-02 until 2020-12) are used in this study.
The official 1-minute BSRN dataset, which covers the whole temporal range and additional variables, is described by \citeA{knap_bsrn_2022}.
Direct irradiance is measured by a CH-1 pyrheliometer, which has a response time of 95\% in 7 seconds and 99\% in 10 seconds \cite{kipp-ch1}.
Diffuse and global irradiance are measured by the CM-22 pyranometer, which is slightly faster \cite{kipp-cm22}.
This means that while the sampling rate is 1 Hz, the resolved resolution is slightly lower, the implications of which are discussed in Section \ref{sec:powerlaw}.
Quality control is optimised for the 1 Hz dataset in this study.
Anomalous data, e.g. due to maintenance, is filtered out, but decorrelation at short time scales between the instruments, because they are a few meters apart, is kept in. 
Solar elevation angles ($\alpha$) are calculated using PySolar \cite{pysolar} at a 1 minute resolution for Cabauw (51.968 N, 4.927 E), and linearly interpolated down to 1 second. 
Direct horizontal irradiance is calculated by multiplying direct irradiance with $sin(\alpha)$.
CAMS McClear version 3.5 \cite{gschwind_improving_2019} is our source dataset for clear sky irradiance (GHI$_\mathrm{cs}$), which takes into account the effect of aerosols and atmospheric gases.
This is a model, rather than an observation, though similar to solar elevation angles it provides essential context for irradiance measurements, discussed further in Section \ref{sec:methodclass}.
It is available at a 1 minute resolution, which we linearly interpolate to 1 second for direct comparison to observations.
An example time series of the irradiance data is shown in Figure \ref{fig:example_timeseries}.

Meteorological context for the irradiance measurements is provided by the following datasets.
From the Cabauw site, wind measurements at 200 meters are publicly available at a 10 minute resolution for our selected period of 2011 until 2020 \cite{knmitower}.
The CLAAS2 satellite product \cite{benas_msg-seviri-based_2017} provides cloud cover, cloud top pressure (CTP), and cloud optical thickness (COT), available every 15 minutes during day time at an approximate spatial resolution of 20 km$^2$ over Cabauw.
Cloud top pressure and optical thickness from this product were validated against various other measurement techniques, by which the authors confirmed the reliability of the satellite-derived cloud physical properties \cite{benas_msg-seviri-based_2017}. 
Cloud type classifications are based on a simple cloud top pressure and optical thickness categorization (see \citeA{ISCCP_algorithm_h_series} ISCCP algorithm description, their Figure 20).
The 9 types are cumulus (Cu), stratocumulus (Sc), stratus (St) for low clouds; altocumulus (Ac), altostratus (As), and nimbostratus (Ns) for middle clouds; cirrus (Ci), cirrostratus (Cs), and cumulonimbus (Cb) for high clouds. 
Cu, Ac, and Ci are the optically thinnest clouds for each altitude, and St, Ns, and Cb the thickest, where Cb spans from low to high altitude as the only exception in this list.
This 9-type CTP vs. COT classification an oversimplification of reality, and there are alternative ways to set thresholds \cite<e.g>{hahn_isccp_2001}.
However, in this study we only use it to group together clouds conditions of various altitudes and optical thicknesses in a more intuitive way, rather than make strong claims for specific cloud types.
The main limitations are that both the cloud fraction and actual cloud optical thickness contribute to a higher reported optical thickness in a satellite pixel, which is a result of limited spatial resolution, and higher clouds can obscure lower clouds, the implications of which are discussed further in Section \ref{sec:cloudtypes}.

The spatial satellite product is converted to a time series representative for the BSRN station by determining the most common cloud class within a 10 km radius around Cabauw.
Results are similar for 5 or 15 km radii (Figure \ref{fig:ce_vs_cloud_sensitivity}).
A smaller radius is not possible due to satellite resolution (pixel area $\approx$ 20 km$^2$) and larger radii become unrepresentative for Cabauw. 
A few meters next to the BSRN station is a nubiscope \cite{wauben2010laboratory}, which measures cloud cover at a 10 minute interval, and is used here only to visually validate our satellite time series processing.
Cloud type and cover from the satellite, and the nubiscope as a reference, are included in Figure \ref{fig:example_timeseries}.
Although the nubiscope would be a more reliable instrument for getting representative cloud cover for Cabauw, it cannot classify cloud types like the satellite product.

\subsection{Time Series Classification}\label{sec:methodclass}
Deriving cloud enhancement (CE) and cloud shadow statistics starts with classifying all irradiance time series data. 
We define a shadow where direct normal irradiance (DNI) is below 120 W m$^{-2}$, which is the inverse of what the World Meteorological Organization defines as sunshine (DNI $\geq$ 120 W m$^{-2}$, \citeA{WMO_measurement_ch8}).
Cloud enhancement refers to a situation where global horizontal irradiance (GHI) exceeds the reference clear-sky (GHI$_\mathrm{cs}$) due to cloud radiative effects, though the exact definition and interpretation varies in literature.
\cite{gueymard_cloud_2017} discusses various such definitions, and favours fixed values such as 1000 W m$^{-2}$ or a fraction of extraterrestrial irradiance as a reference, though this is within the context of photovoltaic applications.
In our study, however, we aim to study the effect of clouds, and therefore we use a reference model (CAMS McClear) that calculates clear-sky irradiance for a geographical location, including the effect of aerosols and atmospheric composition such as water vapour and ozone.
Using a model introduces some uncertainty, and in reality observed GHI still fluctuates slightly in cloud-free conditions (see e.g. 14-15 UTC in Figure \ref{fig:example_timeseries}), both of which make the detection of cloud enhancement tricky for the weakest cases.
To prevent false positives in the detection of cloud enhancements, we first apply an activation threshold of GHI exceeding GHI$_\mathrm{cs}$ by 1\% and 10 W m$^{-2}$. 
Both a relative and absolute threshold are used to deal with high and low solar zenith angles, where 10 W m$^{-2}$ is based on 1\% of the typical order of magnitude for clear-sky irradiance around noon for Cabauw.
When the threshold is reached, adjacent measurements are also marked as cloud enhancement so long as they exceed GHI$_\mathrm{cs}$ by 0.1\%.
Edge cases at low solar elevation angles are removed by requiring DNI to be at least 10 W m$^{-2}$.
All of these thresholds are subjective to some extent, but chosen to enable us to capture all but the weakest of cloud enhancements.
The residual third class is simply 'sunshine', and an example of all three are illustrated in Figure \ref{fig:example_timeseries}.

Within the classified irradiance time series, we call sections of cloud enhancements or shadows 'events'. 
For every event, the start and end time are used to subset radiation and meteorological data, such that every cloud enhancement and shadow event can be characterized.
These include statistics of event duration, maximum cloud enhancement strength 'max(CE)', minimum direct irradiance 'min(DNI)', mean 200 meter wind speed, dominant cloud type, maximum cloud top height, and mean solar elevation angle.
Event statistics such as these allows for filtering of events according to additional criteria, e.g. for comparing events of different magnitudes.
Events at the start or end of daylight are ignored in the analyses, because it is not possible to determine their true duration and other event statistics.
Only events with a solar elevation angle above 5 degrees are included, to filter out small absolute errors in irradiance observations that become relatively big at sunrise or sunset.

\subsection{Distribution Fitting and Conceptual Model}\label{sec:fitmodel}
A common way to analyse irradiance variability is by finding a best fit for the distribution of certain quantities of variability, which in this work are the spatiotemporal scales of cloud shadows and enhancements.
The probability density function of cloud shadow sizes we present in Section \ref{sec:dists}b is analysed with the help of the Powerlaw package \cite{alstott_powerlaw_2014}, which takes care of the non-trivial fitting of heavy-tailed distributions. 
For a power law f($x$) = $x^{\alpha}$e$^{x \lambda}$ this package automatically finds the start of the power law scaling range, $x_{min}$, where $x$ is the shadow size and f($x$) is the according probability density.
It does so by creating power law fits starting from every value possible for $x_{min}$, and finding the one that has the smallest Kolmogorov-Smirnov distance between the data and the fit \cite{alstott_powerlaw_2014}. 
An optional upper end of the scaling range, $x_{max}$, can be manually set to the highest possible expected value or a theoretical limit.
In the case of this study, the maximum expected shadow size is the duration of daylight with a solar elevation angle above 5 degrees multiplied by the mean 200 meter wind speed for that part of the day.
We find this to be 296 km, which is an average over all months in 2012 until 2020.
Similarly, shadow sizes approaching the observational limit $x_{max}$ are undersampled, because they are ignored when they overlap with sunset or sunrise as mentioned in the previous section.
The exponential term e$^{x \lambda}$ in the power law f($x$) corrects for effect. 
The same procedure applies to summer (May to August) and winter (November to February) subsets of the data for sensitivity analyses.
Additional sensitivity tests (yearly variability, solar elevation angle) are done using the [$x_{min}$, $x_{max}$] range found for the whole dataset, because the reduction in amount of samples leads to a noisier distribution and less robust values for $x_{min}$.

\subsection{Numerical modelling}\label{sec:nummodel}
The large-eddy simulations (LES) are performed with MicroHH \cite{van_heerwaarden_microhh_2017}. 
Several new physics options were recently implemented to allow simulations with the complexity of realistic weather: the RTE-RRTMGP radiative transfer solver which calculates radiative transfer in 2D for each model column \cite{pincus2019}, an interactive land-surface model closely following HTESSEL \cite{balsamo2009}, and a single moment ice microphysics scheme \cite{tomita2008}.
In order to facilitate LES simulations of real-life weather, MicroHH was initialised and coupled to ERA5 \cite{hersbach2020} using a similar method as described by \citeA{neggers2012, schalkwijk2015, heinze2017}.
In this coupling the atmosphere and soil are initialized from ERA5, and the influence of spatial and temporal variability in the large-scale weather are imposed on the LES as time and height varying external forcings.
These forcings include the large-scale advective tendencies of temperature, moisture, and momentum, the geostrophic wind components, and the subsidence velocity.

Three different LES experiments are performed over Cabauw, running from 01-08-2016 00 UTC to 01-09-2016 00 UTC, with horizontal domain sizes (grid spacing $\Delta x$) of 12.8 km ($\Delta x$ = 50 m), 25.6 km ($\Delta x$ = 100 m), and 51.2 km ($\Delta x$ = 100 m).
All simulations use the same stretched vertical grid with 192 levels, starting with a vertical grid spacing of $\Delta z$ = 20 m at the surface, and a total vertical extent of ~18.1 km.

\begin{figure}[ht]
\centering
\includegraphics[width=0.5\textwidth]{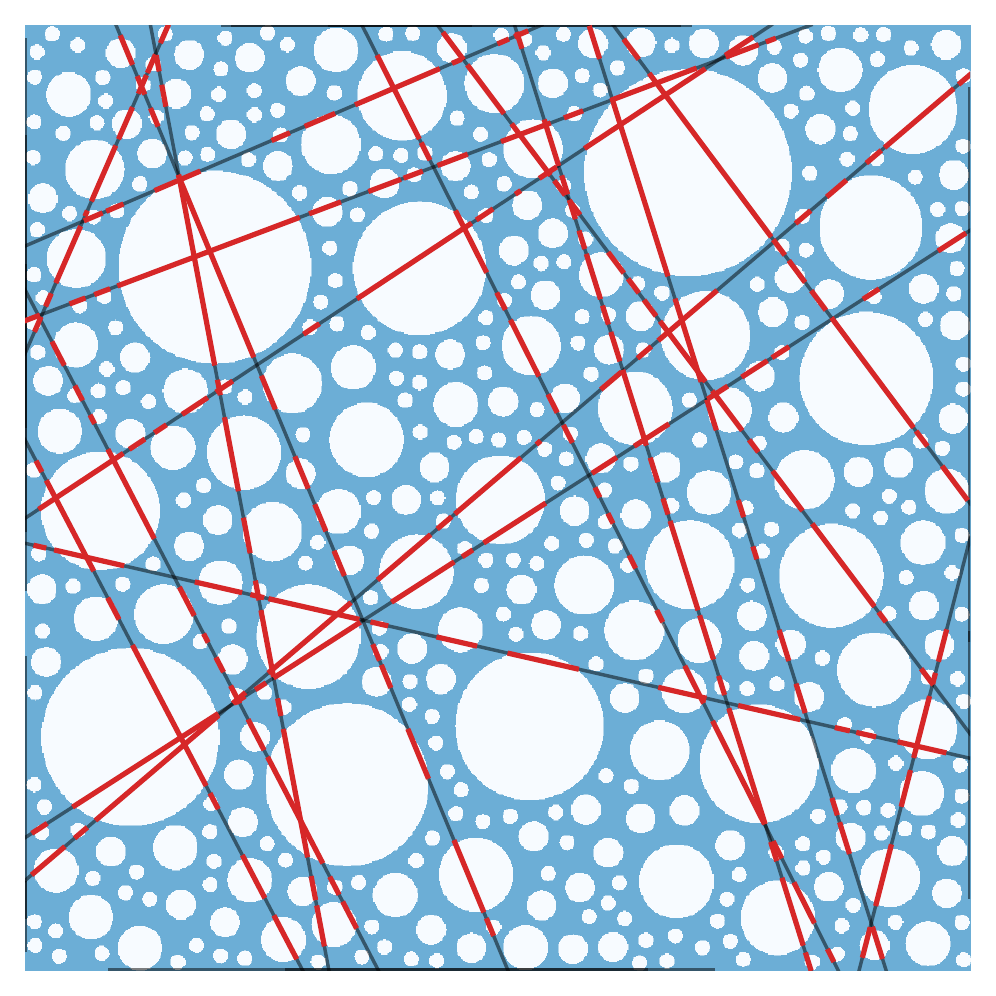}
\caption{\textbf{Top view of an idealised cloud field and the random line sampling technique.} This example contains 765 'clouds', has a 'cloud' cover of 53.6\%, and features horizontal grey lines that represent random sampling transacts with the red sections the measured 'cloud' chord lengths. The size distribution is described by a power law, horizontal units are arbitrary.}
\label{fig:ideal_circles}
\end{figure}

The shadow lengths from LES are obtained using the surface irradiance extracted from an individual model column, sampled at a 5 second frequency.
Shadows are determined using the same definition as used for the BSRN observations, i.e. DNI $\leq$ 120 W m$^{-2}$, with an additional GHI$_\mathrm{cs}$ $>$ 10 W m$^{-2}$ constraint to filter out nights.
The cloud size distributions from the LES simulations are obtained using the cloud tracking method of \citeA{heus_automated_2013}, which identifies individual clouds as spatially continuous areas with a liquid water path (LWP) over a threshold of 5 g kg$\mathrm{^{-1}}$.
From these 2D cloud masks, the cloud sizes are either determined as $\sqrt{A}$ of each cloud area $A$, or by randomly sampling lines through each 2D cloud mask, determining the cloud sizes as the mask-line intersection lengths (or 'chord lengths').
The same methodology is used for the idealised circle model, only with the realistic 2D cloud masks replaced by a numerically generated 'cloud' field consisting of circles.
This field is generated by randomly placing 10$^{5}$ circles, sampled from a power law distribution with a -2.7 slope, inside a domain of 500x500 km$^2$.
The circles are placed ensuring that the individual circles do not overlap, resulting in a circle cover of 40.7\%.
The resulting field is finally analysed using the same methods as used for the realistic 2D cloud masks from LES, with the circle sizes determined from either the circle area, or the intersection between the circles and randomly drawn lines.
Figure \ref{fig:ideal_circles} shows conceptually what such an idealised cloud field would look like, with 765 circles covering 53.6\% of a smaller domain.
The horizontal lines represent random cross sections through this idealised cloud field, mimicking a situation where a cloud fields moves over a single measurement point.
A similar technique for interpreting measurements is applied in \citeA{rodts_size_2003}, we further discuss the implications in our results Section \ref{sec:resmodel}.

\section{Results and Discussion}\label{sec:results}
\subsection{Distributions of Cloud Shadows and Enhancements}\label{sec:dists}
\begin{figure}[ht]
\centering
\includegraphics[width=\textwidth]{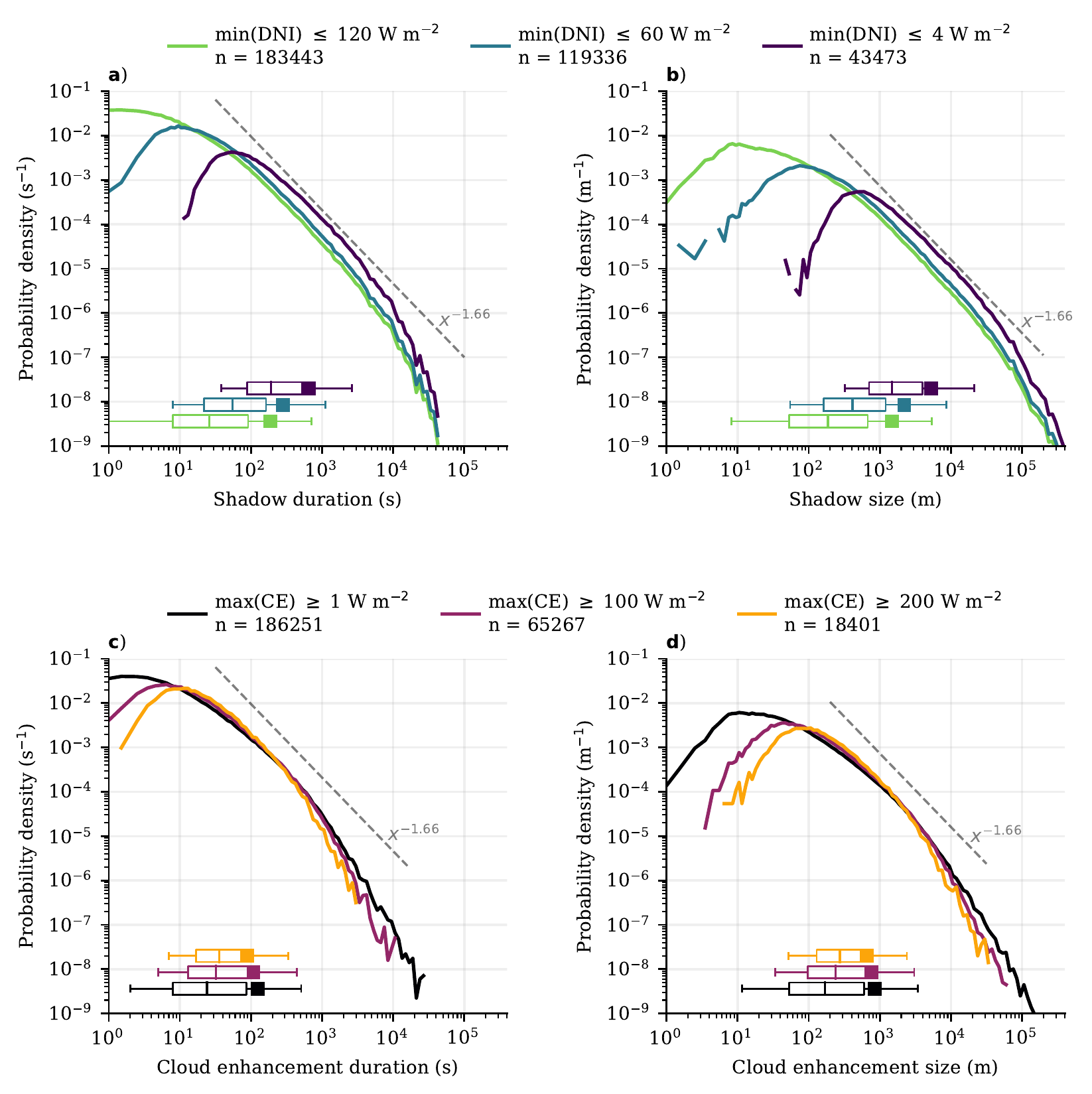}
\caption{\textbf{Duration and size distributions for cloud shadows and enhancements} for three different threshold criteria of minimum or maximum event strength, on a log-log scale. The most strict shadow threshold is set to the event minimum direct normal irradiance 'min(DNI)' $\leq$ 4 W m$^{-2}$, and indicates near complete blocking of direct irradiance by clouds. Cloud enhancement (CE) events are filtered by the maximum absolute clear-sky exceendance 'max(CE)' for an event. Event duration, shown in \textbf{(a, c)} is converted to a length scale using the measured 200 meter wind speed, shown in \textbf{(b, d)}. The gray dashed lines show the cloud size distribution best fit for reference \cite{wood_distribution_2011}. The numbers in the labels are the amount of events that make up the distributions, and the box plots at the bottom span the 5 to 95 percentile, with the mean indicated by a square.}
\label{fig:size_distr}
\end{figure}

Probability density functions for the duration of cloud shadow and enhancement events are illustrated in Figure \ref{fig:size_distr}a,c.
Distributions of 1D horizontal sizes, also known as chord lengths (Figure \ref{fig:size_distr}b,d), are estimated by multiplying event duration by the observed 200 meter wind speed of the tower at the observatory in Cabauw.
This wind speed is an approximation of cloud propagation speed (discussed further in Section \ref{sec:expsense}).
The distributions show that cloud shadows and enhancements cover a wide spatiotemporal range, such that there is not one characteristic time or length scale.
This behaviour is consistent among different levels of strictness in the classification criteria. 
'Dark' shadow events (minimum DNI $\leq$ 4 W m$^{-2}$, i.e. where clouds block all direct irradiance), make up the right tail of the distribution, meaning they result from larger, optically thick clouds.
Conversely, the 10\% strongest cloud enhancement events (maximum CE $\geq$ 200 W m$^{-2}$) are confined to a more narrow range of scales compared to the weaker ones, though they still cover two orders of magnitude, approximately 5 to 500 seconds and 50 to 5000 meters for the middle 90\% of data.
These numbers are in the same range as an analysis using spatial data by \cite{jarvela_characteristics_2020}, except they found shorter duration and size for the strongest peaks in their dataset, possibly explained by the fast response time of their pyranometers compared to ours.
The response time of the instrumentation of the BSRN station we use is in the order of seconds, thus we expect the probability densities in Figure \ref{fig:size_distr} to be negatively biased below 10 seconds (discussed further in Section \ref{sec:powerlaw}).

Shadows and cloud enhancements occur as frequently and over the same spatiotemporal range for the least strict threshold criteria, 18.4 $\pm$ 1.4 and 18.8 $\pm$ 1.6 thousand times per year respectively.
This equals to $\approx$ 50 shadows and cloud enhancements per day, though it will greatly depend on the weather for that day.
Shadows and cloud enhancement occurrence being balanced is perhaps not a coincidence, a cloud that casts a shadow also scatters light next to the shadow, resulting in irradiance enhancement.
But optically thin clouds like scattered cirrus can create weak enhancement without casting shadows, and a cumulus field underneath cirrus will create shadows but possibly no enhancement due to too much attenuated direct irradiance.
There are seasonal differences, where from May to August cloud enhancement events are 10\% more common than shadows, compensated by the winter months such that their total counts converges to within 2\%. 
The scales over which dark shadows (min(DNI) $\leq$ 4 W m$^{-2}$) occur, however, are about a factor 10 larger than that of strong cloud enhancements (max(CE) $\geq$ 200 W m$^{-2}$, see box plots in Figure \ref{fig:size_distr}).
\citeA{gu_cloud_2001} found this asymmetry too, and reported shadows lasting 3 times longer than cloud enhancemen. 
Their findings are based on two months of 1-minute resolution observations, which might explain why the mean duration is 5 times longer than what we find.
Both their and our results suggest there is an asymmetry in spatiotemporal scales between peaks and shadows.
However, the ratio will depend on subjectively chosen thresholds, and also requires high enough resolution to fully resolve the spatiotemporal scales, and is thus not a general rule.
Rather, the true nature of the relationship between the spatiotemporal scales of cloud shadows and enhancements are fully represented by the distributions such as those presented in Figure \ref{fig:size_distr}.

\clearpage

\subsection{Cloud Types Driving Variability}\label{sec:cloudtypes}
Thus far, we have considered irradiance variability by looking at time and length scales of cloud shadows and enhancement.
Optical properties of clouds vary greatly, though, and it matters for irradiance variability whether clouds are isolated, close together, or form a continuous cover.
We use a geostationary satellite dataset \cite<MSG SEVIRI,>{benas_msg-seviri-based_2017} in the following analysis, and combine this with the irradiance dataset by extracting a time series from the spatial data for an area around Cabauw.

\begin{figure}[ht]
\centering
\includegraphics[width=1\textwidth]{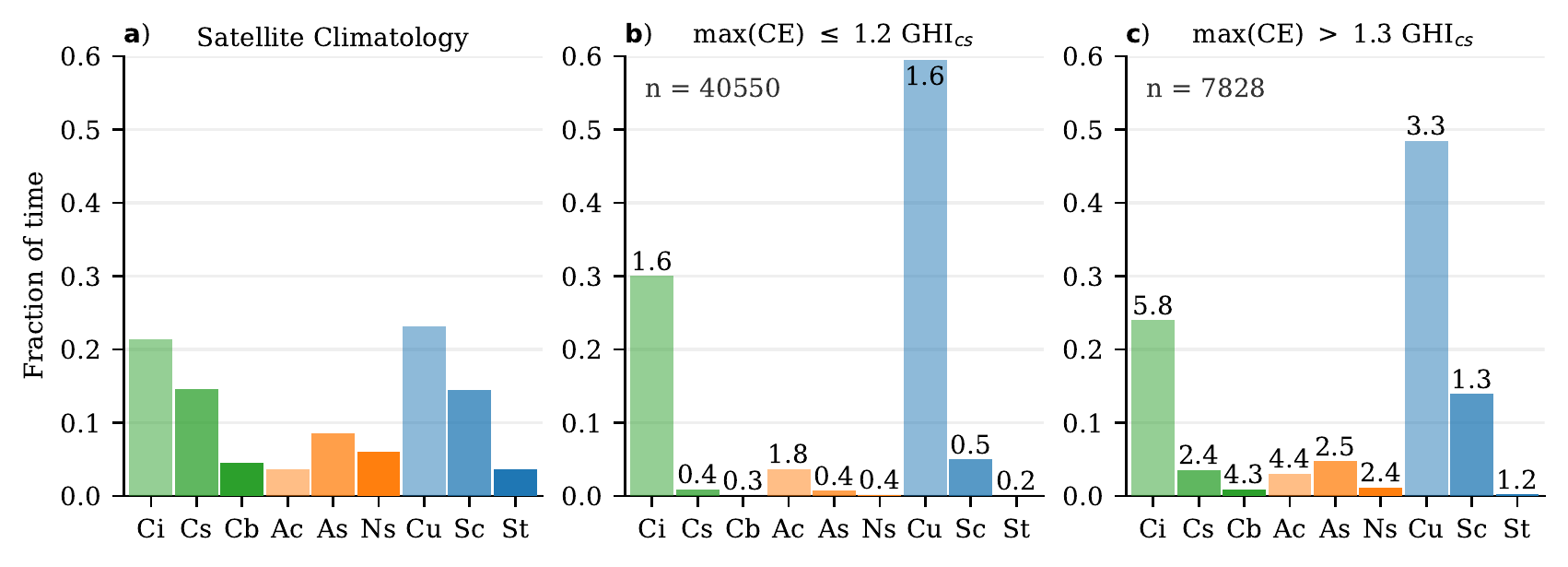}
\caption{\textbf{Cloud types for weak and strong cloud enhancement} based on combining satellite observations \cite{benas_msg-seviri-based_2017} with irradiance observations for Cabauw, the Netherlands. The figure is based on the dominant cloud type in a 10 km radius around the observational site, for the years 2014, 2015, and 2016. \textbf{(a)} shows the cloud type climatology, as probability density under cloudy conditions. \textbf{(b, c)} show the same, but for cloud enhancement events that peak up to 20\% of clear-sky irradiance or above 30\%, respectively. The numbers in the top left indicate the amount of events that make up the distributions. Numbers on the bars indicate mean event duration in minutes. Cloud classes, from left to right, are Cirrus, Cirrostratus, Cumulonimbus, Altocumulus, Altostratus, Nimbostratus, Cumulus, Stratocumulus, and Stratus, and are simple classification based on optical depth versus cloud top height \cite{ISCCP_algorithm_h_series}.}
\label{fig:ce_vs_cloud}
\end{figure}

\begin{figure}[htb]
\centering
\includegraphics[width=1\textwidth]{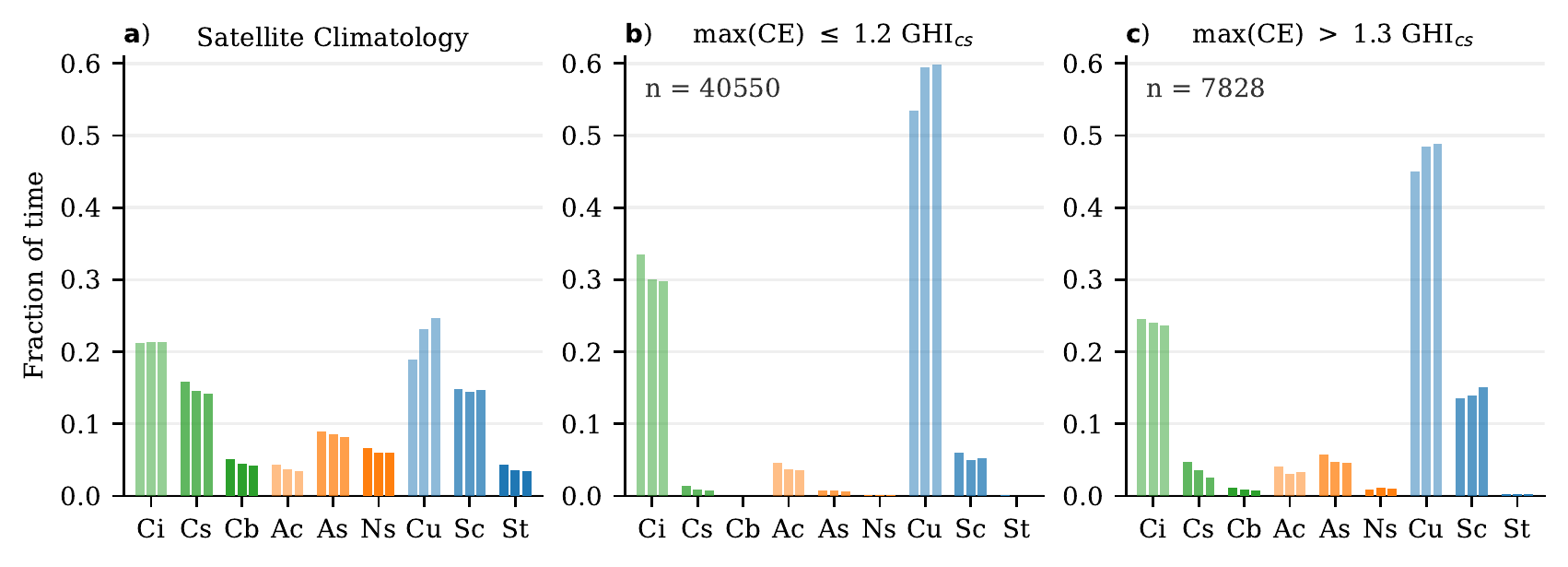}
\caption{\textbf{Cloud types for weak and strong cloud enhancement method sensitivity analysis}. Similar to Figure \ref{fig:ce_vs_cloud}, but here the area radii are 5, 10, and 15 km, plotted from left to right for each cloud type in each subplot.}
\label{fig:ce_vs_cloud_sensitivity}
\end{figure}

Relative to cloud climatology, the conditions under which cloud enhancement occur are dominated by high transparent clouds (cirrus) and broken or scattered low clouds, illustrated in Figure \ref{fig:ce_vs_cloud}.
Cumulus clouds are the main driver of cloud enhancement, both by count and total duration, followed up by cirrus and stratocumulus.
Cirrus is not important for generating variability, though, as it creates relatively long lasting, weak enhancements and is often too optically thin for casting shadows.
Because the cloud classification algorithm uses pixel cloud top height, it can detect high clouds even if cumulus underneath is actually driving irradiance variability, meaning cirrus is likely over represented in this analysis.
Stratocumulus typically has a high cloud cover, but small gaps can cause particularly strong and short cloud enhancements, shown e.g. by \citeA{yordanov_study_2015}, which is likely why it increases in relevance for enhancements exceeding 30\% of clear-sky irradiance (Figure \ref{fig:ce_vs_cloud}c).
Based on our data they generate the shortest lasting cloud enhancement events, averaging roughly one minute versus three minutes for cumulus.
Stratocumulus is in absolute amount of cloud enhancement events almost as common as cumulus for strong cloud enhancement, and together they make up over 70\% of all enhancement events.

\begin{figure}[htb]
\centering
\includegraphics[width=0.75\textwidth]{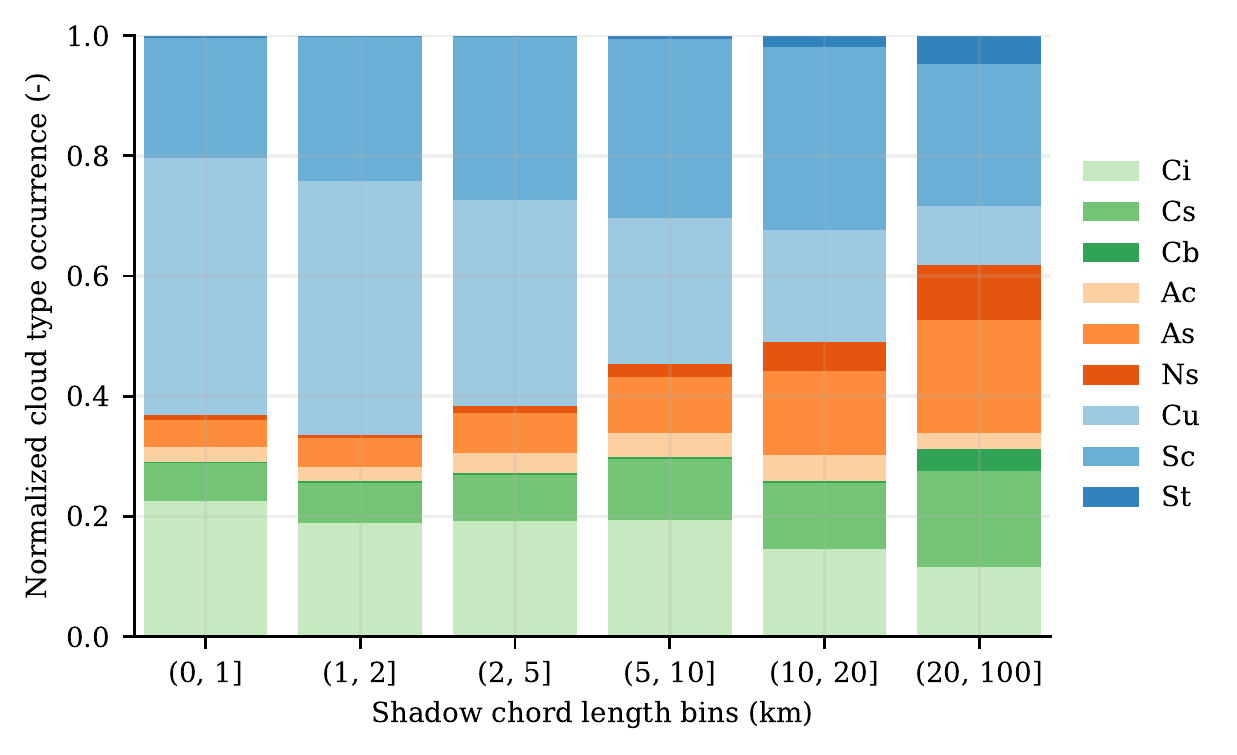}
\caption{\textbf{Dominant cloud type for shadow events of different sizes} for 2014-01 until 2016-12. Histograms are normalized and based on event occurrence count. See Figure \ref{fig:ce_vs_cloud} for the full cloud type names.}
\label{fig:shadow_cloud_classes}
\end{figure}

Results are similar for cloud types causing shadows (Figure \ref{fig:shadow_cloud_classes}), with optically thin and low clouds being the most common until approximately 10 km, and again a likely overestimation of cirrus.
As the shadow length increases, the optically thicker clouds increase in dominance, and the contribution from low clouds decreases.
This means that both peaks and shadows, and thus intra-day irradiance variability, are generated predominantly by boundary layer cloud fields of varying cloud fraction, and make up the majority of the distributions presented in Figure \ref{fig:size_distr}.

\clearpage 

\subsection{Power Laws and Cloud Size Distributions}\label{sec:powerlaw}
\begin{figure}[ht]
\centering
\includegraphics[width=1\textwidth]{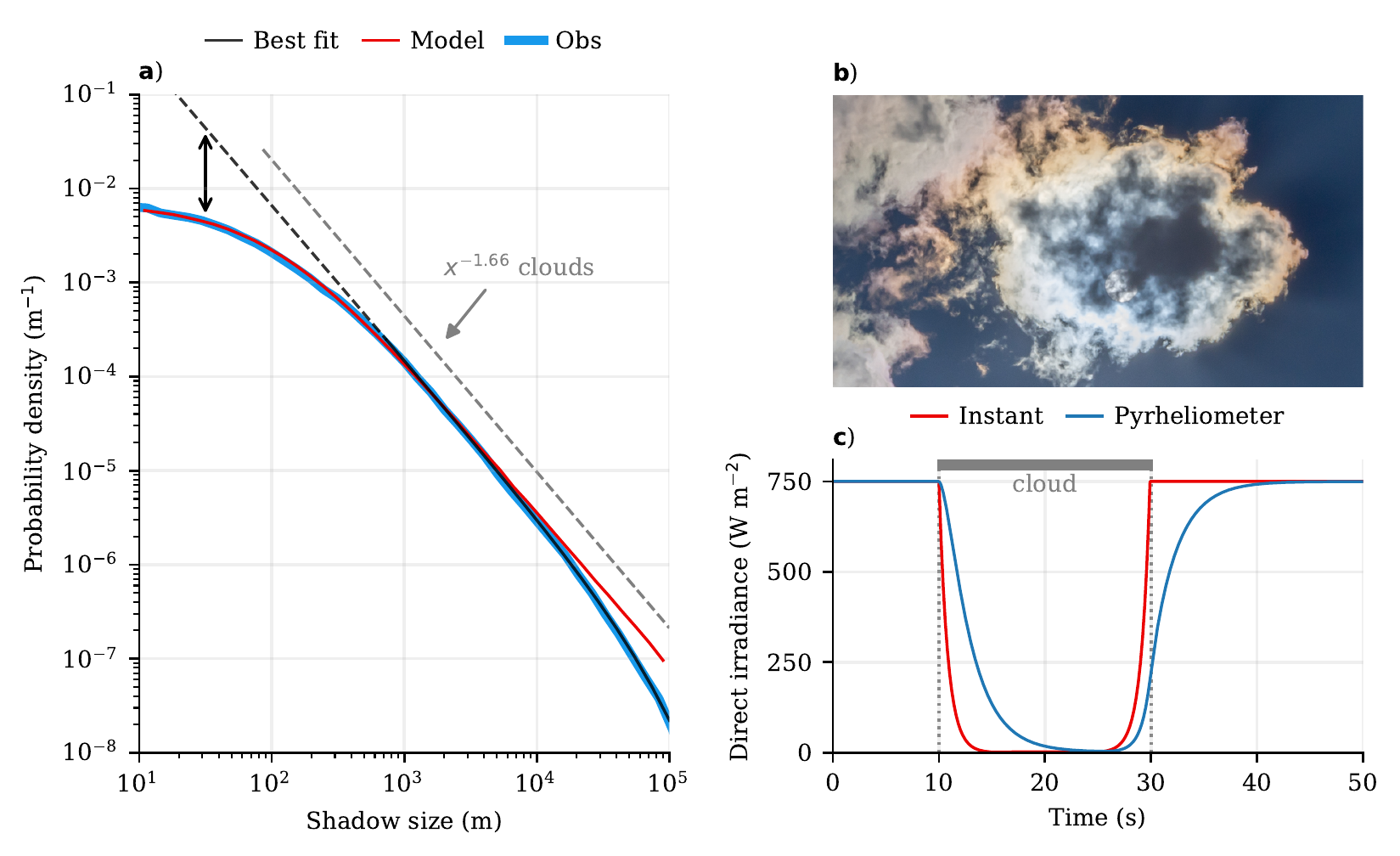}
\caption{\textbf{Shadow size distribution, power law fit, and its relation to cloud size} is illustrated in \textbf{(a)}. The best fit is extrapolated to 10 meters to illustrate the deviation from power law scaling at smaller scales. The observed shadow probability distribution integrates to 1, all other distributions are scaled for visual comparison. The reference cloud size distribution of $x^{-1.66}$ is from \citeA{wood_distribution_2011}. \textbf{(b)} shows a camera picture of a small cloud to illustrate the transparency of cloud edges to direct irradiance. The cloud is $\approx$ 175 meters across (10 Sun diameters and cloud base at 2000 meters). For illustration, the effect of an idealized cloud passage on direct (normal) irradiance versus what a pyrheliometer would measure is schematically shown in \textbf{(c)}. The response of this simulated measurement is exponential, based on the manufacturers specifications \cite{kipp-ch1}ove. The schematic cloud has smooth edges, inspired by mean profiles of liquid water in cumulus \cite{rodts_size_2003} and the picture in \textbf{(b)}.} 
\label{fig:clouds_to_shadow}
\end{figure}

Cloud shadow and enhancement distributions in Figure \ref{fig:size_distr} exhibit power law scaling over multiple orders of magnitude, most notably for shadow sizes.
For the shadow size, a statistical best fit is found for a truncated power law function f($x$) = $x^{\alpha}$e$^{x \lambda}$ for $x \in$ [0.765, 296] km, with parameters $\alpha$ = -1.643 $\pm$ 0.039 and $\lambda^{-1}$ = 79.4 $\pm$ 13.1 km (Figure \ref{fig:clouds_to_shadow}b).
The exponential truncation term e$^{x \lambda}$ corrects for the undersampling of the scales that approach the observational limit ($\approx$ 296 km, Section \ref{sec:fitmodel}).
The uncertainty in $\alpha$ ($\pm$ 0.039) is the year-to-year best fit standard deviation.
Our best fit for exponent $\alpha$ = -1.643 falls within the range of the power law exponent that describes the cloud size distribution $\beta$ = -1.66 $\pm$ 0.04 as reported by \citeA{wood_distribution_2011}.
Ten years of data and thus a wide variety of cloud types and weather conditions have produced this scaling exponent, and exceeds the scaling ranges reported in previous studies \cite{gu_cloud_2001, tabar_kolmogorov_2014, madhavan_multiresolution_2017}.

Below approximately 765 meters, the observed shadow lengths statistically deviate from the cloud size distribution.
Using a simple model, we can conceptually show how shadows down to 10 meters are still linked to cloud size distribution.
We introduce length scale $L$, which represents the size of cloud edge transparency $L_{ce}$ to direct irradiance (Figure \ref{fig:clouds_to_shadow}b), plus a sensor response time bias $L_{sb}$ (schematically shown in Figure \ref{fig:clouds_to_shadow}c).
The effective reduction of shadow length compared to cloud length is because of lower liquid water content at cloud edges \cite{rodts_size_2003}, making cloud edges more transparent to direct sunlight.
I.e., an optical thickness of $\tau <$ 1.9 for a typical direct normal irradiance of 750 W m$^{-2}$ already results in more than 120 W m$^{-2}$.
The simple model randomly generates a set of 1D cloud sizes from a prescribed cloud size distribution, and subtracts a fixed length scale $L$ from each cloud sample.
Figure \ref{fig:clouds_to_shadow}b shows that the resulting distribution for $L$ = 100 m exactly matches shadow length observations.
The value we find for $L$ may be specific to our location, and is rather a value that works for the total 10 years of cloud conditions combined with solar angles, not one that would work for every individual cloud passage.
Higher typical cloud bases would make the shadow to sunshine transitions more spread out at the surface, and some clouds have much sharper edges than others, which could increase or decrease $L$ respectively.
Perhaps more importantly for the analysis of this dataset, part of the length scale $L$ is a result of  the non-instant response of the thermopile sensor, making $L$ = 100 m an upper limit estimate of the true length scale of irradiance transitions ($L_{ce}$).
There are some techniques that attempt to reconstruct the original true 1 Hz signal from the measured slower signal by deconvolution, such as \cite{ehrlich_reconstruction_2015}.
However, we have no accurate 1 Hz observations to validate against, hence disentangling the contribution of sensor response time bias and cloud edge transparency remains a challenge.
Instead, we make an estimation of the approximate contribution to the length scale $L$ based on manufacturer's specification of sensor response time and typical cloud movement speeds, assuming the measured response to a signal behaves similar to what is illustrated in \ref{fig:clouds_to_shadow}c.
Our pyrheliometer sensor bias is $\approx$ 5 seconds (Section \ref{sec:data}), which for $\approx$ 10 m s$^{-1}$ wind speed (10-year average 200 meter wind speed from the tower) gives $L_{sb}$ $\approx$ 50 meters, which is half of $L$.
The best estimate of the transparency length scale is then $L_{ce} = L - L_{sb}$ $\approx$ 50 meters, or 25 meters for each edge.
Newer generation thermopiles with reduced response times, or fast responding semi-conductor instruments \cite<e.g.>{heusinkveld_new_2022}, can improve upon this estimate by taking away the uncertainty of response time and improve the probability density calculations at the shortest scales in Figure \ref{fig:size_distr}.
Having such accurate measurements can also help research into the cloud size distribution at scales below 100 meters by using the irradiance signal as a measurement technique.
Based on our analysis of the current 1 Hz dataset we nonetheless conclude that the spatiotemporal scales of cloud enhancements and shadows are set by the cloud size distribution, a concept we explore further in the next two sections.

\subsection{Exponent Sensitivity}\label{sec:expsense}
Our exponent $\alpha$ = -1.643 for cloud shadow power law scaling is within a percent of Kolmogorov's scaling exponent of -$5/3$ for isotropic turbulence, which raises the question of whether irradiance variability stems from something fundamental and universal in fluid mechanics.
Although our presented distributions in Figure \ref{fig:size_distr} partially exceed the spatiotemporal range of where Kolmogorov's -$5/3$ scaling is valid, there is evidence supporting an extension of this scaling into the mesoscale \cite{callies_transition_2014}. 
Power law scaling is also found in the atmosphere for a wide range of scales for properties related to clouds, some examples are wind and temperature \cite{nastrom_climatology_1985}, total water variability \cite{schemann_scale_2013}, and cloud mass flux \cite{sakradzija_what_2017}. 

A lot of complexity is hidden in our observed cloud shadow scaling analysis.
Thus far we have taken for granted the complex 3D geometry of clouds, multi layered cloud fields, resulting shadow projections onto the surface as a function of solar elevation angle, decoupled 200 m wind from cloud field propagation velocity, and changing cloud shapes as it passes over the instrument.
Sensitivity analyses reveal this complexity to some extent, and help gain some insight into how generally applicable our results are.
Constructing shadow size distributions for only winter or summer data shows smaller shadows are more common in summer ($\alpha$ = -1.79) than in winter ($\alpha$ = -1.56), potentially driven by frequent summertime cumulus over land for some regions as suggested by \citeA{wood_distribution_2011}.
However, selecting only summer data, and comparing low (5 to 40 degrees) to high ($>$ 45 degrees) solar elevation angles show an effect of similar magnitude.
Using satellite data to select only low clouds, such that the 200 m wind speed is a more reasonable assumption, shows largely unchanged size distributions, likely because the majority of clouds are low (Section \ref{sec:cloudtypes}).
By comparing individual years, we find best fit exponents ranging between -1.57 and -1.70, which indicates variability in cloud size distributions and their timing with solar angles play a role.
Indeed, the driving cloud size distribution power law exponents are known to vary regionally, yearly, seasonally, and per time of day \cite{berg_temporal_2008, wood_distribution_2011, schemann_scale_2013, laar_investigating_2019}, suggesting -$5/3$ scaling in irradiance variability may not be universal.
Power law scaling is nonetheless a robust finding in all our sensitivity tests, and variations on the slope are small enough that the interpretation of its link to cloud size distributions remains unaffected, is likely universal, and finds it origin in the scaling of atmospheric properties.
We draw this general conclusion from a single observational given the apparent robustness of power law scaling under variations in solar position and season, and similar scaling found globally for cloud size distributions in other studies.
Furthermore, there is evidence of power law scaling in irradiance variability for different geographical and climate regions.
Despite different analysis methodologies and limited temporal extent, \cite{tabar_kolmogorov_2014} find -$5/3$ scaling of irradiance variability in the time domain for Hawaii and Germany, and \cite{gu_cloud_2001} estimate a similar scaling with even fewer temporal data just south of the Amazon region in Brazil.

\subsection{Testing Measurement Technique using a Cloud-Resolving Model}\label{sec:resmodel}
There is a rich body of literature on cloud size distributions which we can relate our results to, including the interpretation of power law slopes and measurement technique sensitivity.
One aspect is how a 1D method, i.e. time series or random line transacts through a 2D cloud field, result in different size distributions than a 2D method, such as satellite imagery from which you can derive cloud areas.
Cloud cover derived from a 1D method is equivalent to a 2D method, but the cloud size that contributes most to total cloud cover shifts to smaller scales for 1D methods, as shown by \citeA{rodts_size_2003}.
This has to do with 1D methods underestimating the size of objects, because of the high chance of only slicing part of an irregular shape, but also having a higher chance of detecting a larger object, which \citeA{rodts_size_2003} are able to analytically compensate for.
However, \citeA{berg_temporal_2008} find in their analysis that complicated cloud geometry (shape, tilt, overlap) has a bigger influence than measurement technique.

\begin{figure}[ht]
\centering
\includegraphics[width=1\textwidth]{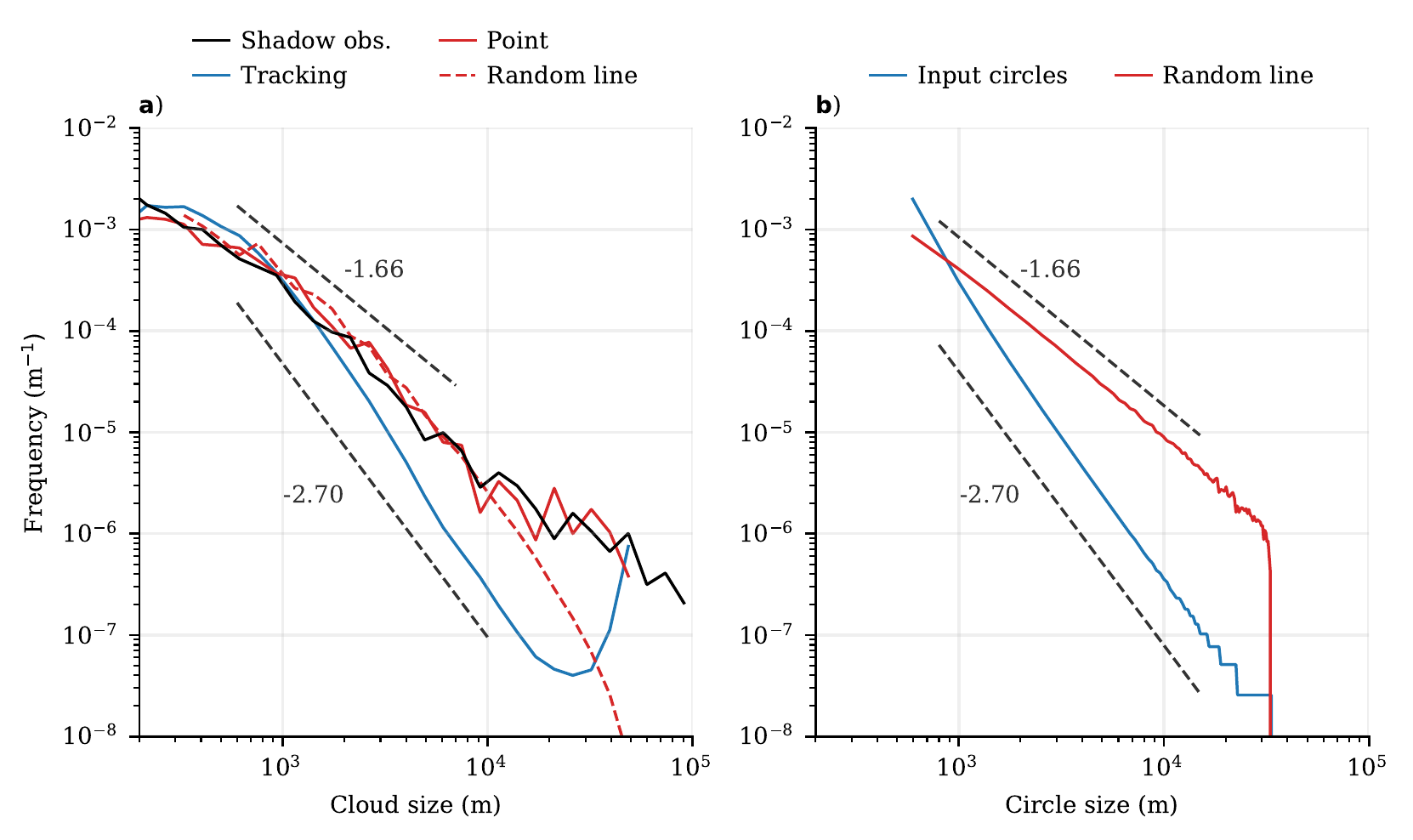}
\caption{\textbf{Different techniques of measuring cloud or shadow length} and their effect on the power law slope. Large-eddy simulation is run for August 2016 for the location of Cabauw, where the irradiance observations are made. The domain size is 51.2 x 51.2 km$^2$ with a horizontal resolution of 100 m. \textbf{(a)} compares observations to point, random line, and cloud tracking techniques based on cloud fields from this model. Gray dashed lines indicate the two power law slopes from 1D and 2D methods over a range where they are largely unaffected by model model resolution and domain size limitations. \textbf{(b)} compares the probably distribution of circle chord lengths using the random line technique applied to a 2D field of circles generated from a power law distribution with a slope of -2.7.}
\label{fig:csd_meaning}
\end{figure}

\begin{figure}[ht]
\centering
\includegraphics[width=1\textwidth]{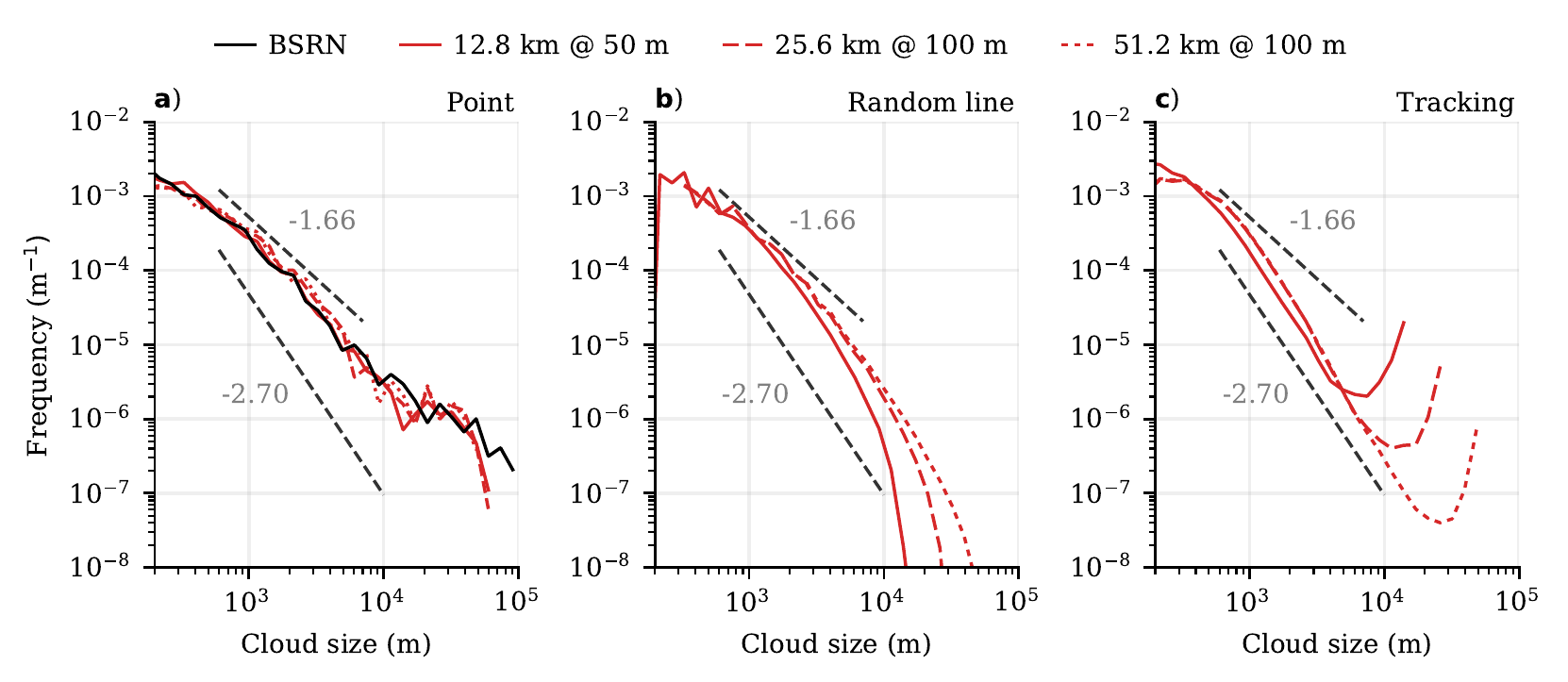}
\caption{\textbf{Different techniques of measuring cloud or shadow length} similar to Figure \ref{fig:csd_meaning}a, but for different domain size and resolution.}
\label{fig:les_csd_sensitivity}
\end{figure}

To test this on our dataset, we have set up a cloud resolving model (LES) to reproduce one month of summer weather conditions at Cabauw, and an idealised circular cloud model similar to \citeA{rodts_size_2003}. 
Figure \ref{fig:csd_meaning}a shows that our model is able to reproduce cloud fields that match the observed distribution, with a power law slope close to -1.66 for 1D methods (random line and simulated point measurements).
Cloud tracking, i.e. finding the areas of all clouds in the domain and converting to 1D length scales by taking the square root, results in a slope close to -2.7, similar to what \citeA{heus_automated_2013} found.
Constructing a 2D field of circles of which the radii distribution is described by a -2.7 power law slope is shown in Figure \ref{fig:csd_meaning}b, and illustrates that if you use a random line sampling technique on this 2D field, the -2.7 slope changes to -1.66, a similar shift in slope as found for cloud resolving model with realistically shaped clouds.
Model domain size or horizontal grid resolution affect the upper and lower end of the resulting distributions, but consistently reproduces this change in slope in a sensitivity analysis (Figure \ref{fig:les_csd_sensitivity}).
A power law slope of -2 means all cloud sizes contribute equally to cloud cover, but for steeper slopes ($<$ -2) this shifts the importance to smaller clouds and vice versa for less steep slopes \cite{wood_distribution_2011}.
This means the observed characterisation of 1D irradiance variability caused by cloud shadows and enhancement (Figure \ref{fig:size_distr}) is driven relatively more by larger clouds, whereas the spatial 2D distribution of irradiance variability is relatively rich in smaller clouds.

\clearpage

\section{Conclusion and Implications}
Based on a ten-year dataset of high resolution observations of surface solar irradiance variability, complemented with modelling and studies on cloud size distributions, we have shown how the cloud size distributions drive the spatiotemporal scales of irradiance variability.
These scales of variability cover multiple orders of magnitude regardless of season, time of day, measurement technique or detection criteria, with the following implications.
To model surface irradiance variability to its full extent and reproduce observed distributions of cloud shadows and enhancements, scales down to seconds or tens of meters ($L_{ce}$ $\approx$ 25 meters) need to be resolved. 
Similarly for observational campaigns, we recommend using fast responding instruments with at least 1 Hz sampling to directly observe the scales down to which clouds occur and contribute to irradiance variability.
Most variability is driven by scattered or broken boundary layer clouds, so focusing on these weather types in observing and modelling efforts is an effective research strategy, both for time series (1D) and spatial (2D) techniques.
Already, though, our presented distribution shapes and how they originate from cloud size distributions provide observational guidance towards an accurate representation of irradiance variability in models.
These results thus have implications for weather and climate in general, but also solar energy and the photosynthetic response of vegetation under variable light, where detailed knowledge of surface solar irradiance is essential.

\clearpage

\section{Open Research}
All data, code and software used for this research is publicly available.
Surface solar irradiance observations of Cabauw are available on Zenodo \cite{bsrn_1hz_p1}, 
Time series classifications, supplementary data, quality control, event statistics and satellite data are published in a separate dataset on Zenodo \cite{bsrn_1hz_p2}.
Satellite data for an area around Cabauw is taken from the CLAAS2 open access dataset described in \citeA{benas_msg-seviri-based_2017}, and included in the previous dataset for ease of use.
The nubiscope data \cite{wauben2010laboratory} is available at the KNMI Data Platform: \url{https://dataplatform.knmi.nl/dataset/cesar-nubiscope-cldcov-la1-t10-v1-0}.
All code to reproduce the classifications from the basic irradiance observations, event statistics, and figures presented in this paper is  archived at \url{https://zenodo.org/record/7472545}.

The model used for large-eddy simulation, MicroHH, input files, output statistics and idealized circle model are all archived at \url{https://zenodo.org/record/7100335}.

\acknowledgments
C.v.H, W.M., and B.v.S. acknowledge funding from the Dutch Research Council (NWO) (grant: VI.Vidi.192.068).
B.v.S. acknowledges funding from NWO (grant 184.034.015). 
The large-eddy simulations were carried out on the Dutch national e-infrastructure with the support of SURF Cooperative.
Our thanks goes out to Jordi Vilà and Frank Kreuwel for providing insightful feedback on an earlier version of the manuscript.



%
%



\bibliography{zotero_library.bib}

%
%
%
%
%

\end{document}